\documentclass[10pt]{iopart}

\usepackage{iopams}
\usepackage{amssymb}
\usepackage{amsfonts}
\usepackage{mathrsfs}
\usepackage{bbm,bm}
\usepackage{graphicx}
\usepackage{xcolor}
\usepackage{hyperref}
\hypersetup{colorlinks=true,breaklinks=true,pdfstartview=FitH,linkcolor=blue,citecolor=blue}
\usepackage{cite}

\newcommand{\bra}[1]{\langle #1\vert}
\newcommand{\ket}[1]{\vert #1\rangle}

\newcommand{\id}{\mathrm{id}}
\newcommand{\diag}{\mathrm{diag}}%

\newtheorem{definition}{Definition}

\newtheorem{proposition}{Proposition}
\newtheorem{theorem}{Theorem}

\begin{document}

\title{Quantum discord of two-qubit rank-two states}

\author{Mingjun Shi$^1$, Wei Yang$^1$, Fengjian Jiang$^1$ and Jiangfeng Du$^{1,2}$}

\address{$^1$ Department of Modern Physics, University of Science and
Technology of China, Hefei, Anhui 230026, People's Republic of China}
\address{$^2$ Hefei National Laboratory for Physical Sciences at Microscale \&
Department of Modern Physics, University of Science and Technology of China,
Hefei, Anhui 230026, People's Republic of China}

\ead{\mailto{shmj@ustc.edu.cn}}
\ead{\mailto{djf@ustc.edu.cn}}

\begin{abstract}
Among various definitions of quantum correlations, quantum discord has attracted considerable attention.
The connection between the quantum discord and the entanglement of formation is described by Koashi-Winter relation.
We investigate this relation from the viewpoint of the quantum channel that is isomorphic to the given state.
It is shown that in the case of two-qubit states the channel, on the one hand, determines the form of the quantum steering ellipsoid of the given state, and on the other hand, is closely related to the concurrence of the complement state of the given state.
We also point out that, for two-qubit rank-two state, von Neumann measurement is the optimal choice to achieve the quantum discord. However, for some two-qubit states with the rank larger than two, the three-element POVM measurement is optimal.
\end{abstract}


\maketitle

\section{Introduction}

Quantum correlations between parties of a bipartite quantum system have been the significant and interesting subject in quantum information processing as well as in quantum theory.
There are various measures to quantify the quantum correlations, among which quantum discord has attracted considerable attention.

Quantum discord is originally defined as the the difference between two classically identical (but quantumly distinct) formulas that measure the amount of mutual information of a pair of quantum systems \cite{Ollivier.PhysRevLett.88.017901.2001}.
An alternative way to define quantum discord is by writing it as the difference between total correlation and classical correlation \cite{Henderson.JPhysA.34.6899.2001}. The total correlation is usually quantified by the mutual information $\mathcal{I}$,
\begin{equation*}
    \mathcal{I}=S(\rho^A)+S(\rho^B)-S(\rho^{AB}),
\end{equation*}
where $\rho^{AB}$ is the state of a bipartite quantum system composed of particle $A$ and particle $B$, the $\rho^{A(B)}=\Tr_{B(A)}(\rho^{AB})$ is the reduced state of $A$ ($B$), and $S(\rho)$ is the von Neumann entropy of a quantum state $\rho$.
As for the classical correlation, we adopt the definition given by Henderson and Vedral \cite{Henderson.JPhysA.34.6899.2001}.
Suppose that a positive operator valued measure (POVM) measurement is performed on particle $A$.
The set of POVM elements is denoted by $\mathcal{M}=\{M_k\}$ with $M_k\geqslant 0$ and $\sum_kM_k=\mathbbm1$.
The operation element $M_k$ will produce particular outcome, labeled by $k$, with the probability $p_k=\Tr[\rho^{AB}(M_k\otimes\mathbbm1)]$.
When the outcome $k$ occurs, the postmeasurement state of particle $B$ is $\rho_k^B=\Tr_A[\rho^{AB}(M_k\otimes\mathbbm1)]/p_k$.
That is, the POVM measurement on $A$ will ``realize'' a postmeasurement ensemble $\{p_k,\rho_k^B\}$ for particle $B$.
In other words, the reduced state $\rho^B$ is decomposed into $\sum_k p_k \rho_k^B$ due to the measurement $\mathcal{M}$ performed on $A$.
The amount of information acquired about particle $B$ is then given by $S(\rho^B)-\sum_kp_kS(\rho_k^B)$,
which depends on measurement $\mathcal{M}$.
This dependence can be removed by doing maximization over all POVM measurements, which gives rise to the definition of classical correlation:
\begin{eqnarray}
    C^{A\to B} & =\max_{\mathcal{M}}
    \Big[S(\rho^B)-\sum_k p_k S(\rho_k^B)\Big] \nonumber \\
    & =S(\rho^B)-\min_{\mathcal{M}}\sum_k p_k S(\rho_k^B).
      \label{def: classical correlation}
\end{eqnarray}
Denote the average entropy $\sum_kp_kS(\rho_k^B)$ as $\overline{S}^{A\to B}$, and the minimal value as $\overline{S}_{\min}^{A\to B}$.
Quantum discord is defined as
\begin{equation} \label{def: discord}
    D^{A\to B}=\mathcal{I}-C^{A\to B}
    =S(\rho^A)+\overline{S}_{\min}^{A\to B}-S(\rho^{AB}).
\end{equation}
The key step, and also the major difficulty, is to work out the minimal average entropy (MAE). So we usually focus on the MAE $\overline{S}^{A\to B}_{\min}$.
We call the POVM measurement on $A$ optimal if such a measurement generates the MAE for $B$. The optimal POVM may not be unique.

Similarly one can define classical correlation $C^{A\leftarrow B}$ and quantum discord $D^{A\leftarrow B}$, corresponding to the case that the measurement is performed on particle $B$ to gain the information about particle $A$.
Generally, the quantum discord and classical correlation defined from $A$ to $B$ may not be equal to that defined from $B$ to $A$, meaning that they are asymmetric.
In this paper, we only consider the direction from $A$ to $B$.
For the sake of clarity, we will omit the superscript ``$A\to B$'' in the expressions of average entropy, classical correlation and quantum discord.

The concept of quantum discord is proposed to describe the quantum correlations which are not limited to entanglement.
There is a fundamental difference between entanglement and discord for mixed states, although they are equivalent for pure states.
A typical example is that there exist non-entangled (or separable) states with nonvanishing discord \cite{Ollivier.PhysRevLett.88.017901.2001}.
Quantum discord has received much attention in studies involving fuzzy measurement \cite{Vedral.PhysRevLett.90.050401.2003}, mixed-state quantum computation speedups \cite{Datta.PhysRevA.75.042310.2007,Datta.PhysRevLett.100.050502.2008},
broadcasting of quantum state \cite{Piani.PhysRevLett.100.090502.2008},
complete positivity of dynamics
\cite{Rodriguez.JPA.41.205301.2008,Shabani.PhysRevLett.102.100402.2009},
complementarity and monogamy relationship between classical and quantum correlations
\cite{Oppenheim.PhysRevA.68.022307.2003,Badziag.PhysRevLett.91.117901.2003,Koashi.PhysRevA.69.022309.2004}, dynamics of discord
\cite{Maziero.PhysRevA.80.044102.2009,Mazzola.PhysRevLett.104.200401.2010},
operational interpretations of quantum discord in terms of state merging
\cite{Madhok.PhysRevA.83.032323.2011,Cavalcanti.PhysRevA.83.032324.2011},
and the relation between discord and entanglement
\cite{Yang.PhysRevLett.95.190501.2005,Cornelio.PhysRevLett.107.020502.2011,
      Streltsov.PhysRevLett.106.160401.2011,Piani.PhysRevLett.106.220403.2011}.

Let us focus our attention on the Koashi-Winter (K-W) relation presented in \cite{Koashi.PhysRevA.69.022309.2004}.
There the authors show that there exists a monogamy relation between the classical correlation of a bipartite state $\rho^{AB}$ and the entanglement of formation (EoF) of the complement $\rho^{BC}$.
By ``complement'', we means that both $\rho^{AB}$ and $\rho^{BC}$ are the reduced states of a tripartite pure state $\rho^{ABC}$, namely, $\rho^{AB}=\Tr_C\rho^{ABC}$ and $\rho^{BC}=\Tr_A\rho^{ABC}$.
Then K-W relation tells us that $C(\rho^{AB})+E(\rho^{BC})=S(\rho^B)$, where $C(\rho^{AB})$ is the classical correlation (from $A$ to $B$) of $\rho^{AB}$ and $E(\rho^{BC})$ is the EoF of $\rho^{BC}$.
By noting the definition \eref{def: classical correlation}, we can rewrite K-W relation as $\overline{S}_{\min}=E(\rho^{BC})$.
Obviously, K-W relation establishes the relationship between the EoF and the classical correlation (or minimal average entropy, or quantum discord) in an extended Hilbert space.
It is remarkable that the particle $B$ is the pivotal point of the relationship.

In this paper, the pivotal role played by particle $B$ is investigated from the viewpoint of quantum channel.
It is well known that there is an isomorphism between quantum states and quantum channel
\cite{Jamiolkowski.RepModPhys.3.275.1972,Choi.LinAlgApp.10.285.1975,
      Kraus.StateEffandOp.1983,Horodecki.PhysRevA.60.1888.1999,Arrighi.311.26.2004}.
Restricted to a two-qubit state $\rho^{AB}$, we can say that, associated with $\rho^{AB}$, there is a channel $\Lambda$ acting upon the qubit $B$ of the Bell state
$\ket{\Phi^{AB}_+}=\frac{1}{\sqrt2}(\ket{0^A}\ket{0^B}+\ket{1^A}\ket{1^B})$.
We further show that that the significance of the quantum channel $\Lambda$ is twofold.
On the one hand, $\Lambda$ determines the quantum steering ellipsoid \cite{Verstraete.diss.2002}, which has been used to discuss the quantum discord in \cite{Shi.NJP.13.073016.2011}.
As such, we present an alternative approach to the quantum steering ellipsoid.
On the other hand, $\Lambda$ is closely related to the concurrence of the complement state $\rho^{BC}$.
In fact, since $\rho^{BC}$ is a rank-two state, its concurrence can be evaluated exactly by means of the methods provided in \cite{Hildebrand.JMP.48.102108.2007,Hellmund.PhysRevA.79.052319.2009}.

Subsequently, we point out that K-W relation implies the following two facts.
(i) For any 2-qubit rank-two state, the quantum discord or classical correlation can be evaluated exactly by using K-W relation.
In this case, we need only to perform POVM measurement with two (rather than three or more) operation elements, i.e., von Neumann measurement, on qubit $A$ to achieve the optimal two-state ensemble for $\rho^B$.
The two states in the optimal ensemble have the identical von Neumann entropy.
In addition, we present the geometric description for the optimal ensemble in terms of the quantum steering ellipsoid.
(ii) For some 2-qubit states with rank larger than two, the von Neumann measurement may not be optimal, while the three-element POVM measurement will give a larger classical correlation or a smaller quantum discord.
This fact has been implied by K-W relation, together with the numerical results obtained in \cite{Hellmund.PhysRevA.79.052319.2009}.

This paper is arranged in the following manner.
In Section \ref{sec: channel and ellipsoid}, we consider the isomorphism between quantum states and quantum channels, and show that for two-qubit states the quantum channel determines the quantum steering ellipsoid.
It is found that the ellipsoid is invariant under any filtering operation performed on a specific qubit.
In Section \ref{sec: channel and concurrence}, a given two-qubit state is purified to a tripartite pure state, and the concrete process is presented to evaluate the concurrence of the complement state.
We find that the quantum channel obtained in Section \ref{sec: channel and ellipsoid} is most relevant to the evaluation of the concurrence.
Section \ref{sec: KW relation} is a discussion about K-W relation.
There we indicate that the von Neumann measurement may not be optimal to give the quantum discord or classical correlation for general two-qubit state.
Section \ref{sec: rank two states} is devoted to exact expressions as well as the geometric picture for the quantum discord and classical correlation of any 2-qubit rank-2 states.
In addition, the conjecture proposed in \cite{Shi.NJP.13.073016.2011} is proved.
Two-qubit states with rank larger than two are discussed in Section \ref{sec: larger rank}.
We show that some known results about the EoF can lead further insight into the quantum discord and classical correlation.


\section{Quantum channel and quantum steering ellipsoid} \label{sec: channel and ellipsoid}
It has been shown that quantum steering ellipsoid is a useful tool to discuss quantum discord and classical correlation \cite{Shi.NJP.13.073016.2011}.
In this section, we present an alternative approach to establish this geometric picture, that is, we will show that the quantum steering ellipsoid eventually boils down to the notation of quantum channel.

\subsection{Quantum channels, one-qubit case}

A quantum channel $\Phi$ is a trace-preserving, completely positive map.
We refer the reader to textbooks (e.g. \cite{Nielsen.QCandQI.2000} and \cite{Bengtsson.GeoQuantState.2006}) for a detailed description of quantum channel.
Let $\rho$ be the density operator on $n$-dimensional Hilbert space.
The action of $\Phi$ on a state $\rho$ can be expressed in the operator-sum representation:
\[
\rho\,\to\,\Phi[\rho]=\sum_iK_i\rho K_i^\dag,
\]
with $K_i$ the Kraus operators and $\sum_iK^\dag_iK_i=\mathbbm1$.
We are able to represent the channel $\Phi$ in the form of superoperator, i.e., a $n^2\times n^2$ matrix: $\Phi=\sum_iK_i\otimes K_i^*$.
With the state $\rho$ represented as a ``long'' vector in $n^2$-dimensional Hilbert space, denoted by $|\rho\rangle\!\rangle$, we can write $\Phi[\rho]$ as $\Phi|\rho\rangle\!\rangle$.
In the following, we consider the simple one-qubit case.

Let $X$ be any operator on the two-dimensional Hilbert space, namely,
\[
X=\left(
\begin{array}{cc}
  x_{00} & x_{01} \\ x_{10} & x_{11}
\end{array}
  \right).
\]
Define the 4-vector (in column form) $\bi{X}$ as $\bi{X}=(x_{00},x_{01},x_{10},x_{11})^T$, where the superscript $T$ means the transpose.
The channel $\Phi$ is now expressed as a $4\times4$ matrix, and the action
on $X$ is given by $\bi{X}\to \Phi\bi{X}$.
On the other hand, the operator $X$ can also be written as $X=\frac{1}{2}\sum_{\mu=0}^{3}x_{\mu}\sigma_{\mu}$, with $\sigma_0$ the $2\times2$ identity matrix $\mathbbm1_2$ and $\sigma_i$ for $i=1,2,3$ the Pauli matrices.
That is, $X$ is expressed by another 4-vector $\bi{x}=(x_0,x_1,x_2,x_3)^T$.
Define a unitary matrix $\Upsilon$, which is from \cite{Verstraete.PhysRevA.64.010101.2001}
\begin{equation}
  \Upsilon=\frac{1}{\sqrt2}\left(
  \begin{array}{cccc}
    1 & 0 & 0 & 1 \\
    0 & 1 & 1 & 0 \\
    0 & i & -i & 0 \\
    1 & 0 & 0 & -1
  \end{array}\right).
\end{equation}
We see that $\bi{x}=\sqrt{2}\Upsilon\bi{X}$.
Then in terms of the 4-vector $\bi{x}$, the action of $\Phi$ is
\begin{equation}
  \bi{x}\to\bi{x}'=\Upsilon\Phi\Upsilon^\dag\bi{x}
    =L_{\Phi}\bi{x},
\end{equation}
where $L_{\Phi}=\Upsilon\Phi\Upsilon^\dag$.

By unitary operations, $L_{\Phi}$ can be expressed in the canonical form \cite{King.IEEE.47.192.2001}:
\begin{equation} \label{canonical form of LPhi}
  L_{\Phi}=\left(
  \begin{array}{cc}
    1 & 0 \\ \vec{\kappa} & \Xi
  \end{array}\right),
\end{equation}
where $\vec{\kappa}=(\kappa_1,\kappa_2,\kappa_3)^T$ is a column vector in three-dimensional real space, and $\Xi$ is a $3\times3$ diagonal matrix, $\Xi=\diag(\xi_1,\xi_2,\xi_3)$.
For one-qubit state
$\rho=\frac{1}{2}\sum_{\mu=0}^{3}x_\mu\sigma_\mu$,
the 4-vector $\bi{x}=(x_0,x_1,x_2,x_3)^T$ is transformed to $\bi{y}=L_\Phi\,\bi{x}$.
In this sense, we call $L_\Phi$ the Bloch representation of the channel $\Phi$.

Notice that $\bi{x}$ is distributed in the Bloch ball (including the surface), namely,
$x_1^2+x_2^2+x_3^2\leqslant 1$, or equivalently,
$\bi{x}^T\eta\bi{x}\geqslant 0$ with $\eta=\diag(1,-1,-1,-1)$.
It follows that $\bi{y}$ is constrained by
\begin{equation} \label{LPhi ellipsoid}
{\bi{y}}^TL_\Phi^{-T}\eta L_\Phi^{-1}\bi{y}\geqslant 0.
\end{equation}
Here it is assumed that $L_\Phi$ is invertible.
It is not difficult to see that (\ref{LPhi ellipsoid}) indicates an ellipsoidal region.
Then the geometric meaning of (\ref{canonical form of LPhi}) is clear: $L_{\Phi}$ transforms the unit sphere $\mathcal{S}^2$ in $\mathbb{R}^3$ (i.e., $x^2+y^2+z^2=1$) to an ellipsoidal surface, given by
\[
  \frac{(x-\kappa_1)^2}{\xi_1^2}+\frac{(y-\kappa_2)^2}{\xi_2^2}
    +\frac{(z-\kappa_3)^2}{\xi_3^2}=1.
\]

\subsection{Quantum channels, two-qubit case}
Now we consider the effects of quantum channels on the two-qubit states.
A 2-qubit state $\rho^{AB}$ can be expressed in Hilbert-Schmidt space as $\rho^{AB}=\frac{1}{4}\sum_{\mu,\nu=0}^{3}R_{\mu\nu}\sigma_{\mu}\otimes\sigma_{\nu}$ with $R_{\mu\nu}=\Tr(\rho^{AB}\sigma_\mu\otimes\sigma_\nu)$.
Define $R$-matrix as $R=(R_{\mu\nu})$.
The matrix $R$ can be also expressed as $R=2\Upsilon(\rho^{AB})^R\Upsilon^T$, where $(\rho^{AB})^R$ is the reshuffling of $\rho^{AB}$ \cite{Verstraete.PhysRevA.64.010101.2001}.
Given a bipartite matrix
\[
X=\sum_{i,j;i',j'} x_{i,j;i',j'}\ket{ij}\bra{i'j'},
\]
the reshuffling of $X$ is defined as
\[
X^R=\sum_{i,j;i',j'} x_{i,i';j,j'}\ket{ij}\bra{i'j'}.
\]

Now suppose that qubits $A$ and $B$ are acted upon by local channels $\Phi^A$ and $\Phi^B$ respectively.
The state $\rho^{AB}$ is mapped to ${\rho'}^{AB}=(\Phi^A\otimes\Phi^B)[\rho^{AB}]$.
Expressing $\Phi^A$ and $\Phi^B$ as $4\times4$ matrices, we have the following formula, which  is essentially from \cite{Bengtsson.GeoQuantState.2006}.
\begin{equation} \label{formula of local channels}
  (\rho^{AB})^R \longrightarrow ({\rho'}^{AB})^R=\Phi^A(\rho^{AB})^R(\Phi^B)^T.
\end{equation}
Expressing it in the Bloch representation, we have
\begin{equation}
  R\,\longrightarrow\,R'=L^{A}R(L^{B})^T,
\end{equation}
where $R'$ is the $R$-matrix of ${\rho'}^{AB}$, $L^A=\Upsilon\Phi^A\Upsilon^\dag$ and $L^B=\Upsilon\Phi^B\Upsilon^\dag$.

\subsection{Quantum steering ellipsoid} \label{subsec: steering ellipsoid}

It is well-known that there is an isomorphism between quantum states and quantum operations, in particular, quantum channels
\cite{Jamiolkowski.RepModPhys.3.275.1972,Choi.LinAlgApp.10.285.1975,
      Kraus.StateEffandOp.1983,Horodecki.PhysRevA.60.1888.1999,Arrighi.311.26.2004}.
In this subsection, we use this isomorphism to derive the equation of quantum steering ellipsoid from the quantum channel.

Given a 2-qubit state $\rho^{AB}$,
we assume that the reduced state $\rho^A=\Tr_B\rho^{AB}$ is invertible.
Define the state $\tilde{\rho}^{AB}$ as
\begin{equation}
  \tilde{\rho}^{AB}=\frac{1}{\sqrt{2\rho^A}}\,\rho^{AB}\,\frac{1}{\sqrt{2\rho^A}}.
\end{equation}
Then $\tilde{\rho}^A=\Tr_B\tilde{\rho}^{AB}=\frac{1}{2}\mathbbm1_2$.
The isomorphism between state and channel is expressed as
\begin{equation} \label{rhowan and Pplus}
  \tilde{\rho}^{AB}=(\id\otimes\Lambda)[P_+],
\end{equation}
where $\id$ denotes identity operator and $P_+=\ket{\Phi_+}\bra{\Phi_+}$ with $\ket{\Phi_+}=\frac{1}{2}(\ket{00}+\ket{11})$ a Bell state.

According to formula \eref{formula of local channels}, we have $(\tilde{\rho}^{AB})^R=P_+^R\Lambda^T$.
By noting that $P_+^R=\frac{1}{2}\mathbbm1_{4}$ with $\mathbbm1_{4}$ the $4\times4$ identity matrix, we see that $\Lambda^T=2(\tilde{\rho}^{AB})^R$.
In the Bloch representation, the channel is represented as
\begin{eqnarray} \label{L Lambda}
  L_\Lambda & =\Upsilon\Lambda\Upsilon^\dag
    =2\Upsilon[(\tilde{\rho}^{AB})^R]^T\Upsilon^\dag \nonumber \\
    & =2[\Upsilon^*\Upsilon^\dag\Upsilon(\tilde{\rho}^{AB})^R\Upsilon^T]^T
      =\tilde{R}^TR_+^T,
\end{eqnarray}
where $\tilde{R}$ is the $R$-matrix of $\tilde{\rho}^{AB}$ and $R_+$ is the $R$-matrix of $P_+$, that is, $R_+=\diag(1,1,-1,1)=\Upsilon^*\Upsilon^\dag$.

According to (\ref{LPhi ellipsoid}), the output states of the channel $\Lambda$ are constrained in an ellipsoidal region, which is given by
\begin{equation} \label{ellipsoid in tilde R}
  \bi{y}^T\,\big[L_\Lambda^{-T}\eta L_\Lambda^{-1}\big]\,\bi{y}
    =\bi{y}^T\,\big[\tilde{R}^{-1}\eta\tilde{R}^{-T}\big]\,\bi{y}\geqslant 0.
\end{equation}
Here we assume that $L_\Lambda$ (or $\tilde{R}$) is invertible.
Define the matrix $\tilde{\Omega}=\tilde{R}^T\eta\tilde{R}$.
It is the matrix $\tilde{\Omega}^{-1}$ that determines the ellipsoid,
which we denote by $\tilde{\frak{E}}$.

Define a similar matrix $\Omega=R^T\eta R$ with $R$ being the $R$-matrix of $\rho^{AB}$, i.e., $R=2\Upsilon(\rho^{AB})^R\Upsilon^T$.
An ellipsoid $\frak{E}$ can be defined from $\Omega^{-1}$. The equation is given by
\begin{equation} \label{ellipsoid E}
  \bi{y}^T\,\big[R^{-1}\eta R^{-T}\big]\,\bi{y}=0.
\end{equation}
We will show that $\frak{E}$ is identical to $\tilde{\frak{E}}$.

In fact, the ellipsoid $\frak{E}$ (or $\tilde{\frak{E}}$) is invariant under any local filtering operation performed on qubit $A$.
To see this, let us consider how the matrix $\Omega$ is transformed under the local filtering operation.
The process of any local filtering on $A$ is given by
\begin{equation} \label{LF on A}
\rho^{AB}\longrightarrow(F\otimes\mathbbm1)\rho^{AB}(F^\dag\otimes\mathbbm1),
\end{equation}
where $F$ is a $2\times2$ nonsingular matrix with $F^\dag F\leqslant\mathbbm1$.
The right-hand side of (\ref{LF on A}) is an unnormalized state.
By defining $\Phi_F=F\otimes F^*$, we rewrite (\ref{LF on A}) as
\begin{equation} \label{rhoABR to LF rhoABR}
  (\rho^{AB})^R\longrightarrow \Phi_{F}(\rho^{AB})^R.
\end{equation}
Note that $\Phi_F$ is not a channel, since it is completely positive but not a trace-preserving map.
In the Bloch representation, (\ref{rhoABR to LF rhoABR}) is equivalent to
$R\longrightarrow L_FR$ with $L_F=\Upsilon\Phi_F\Upsilon^\dag$.
By noting that $L_F^T\eta L_F=|\det(F)|^2\eta$, we have
\[
  R^T\eta R\longrightarrow R^TL_F^T\eta L_FR=|\det(F)|^2R^T\eta R.
\]
That is, $\Omega\to|\det(F)|^2\Omega$ for any local filtering operator $F$.
It follows that the ellipsoid $\frak{E}$ is invariant under the operation
$F\otimes\mathbbm1$.

It is remarkable that (\ref{ellipsoid E}) is just the equation of the quantum steering ellipsoid which we introduce in \cite{Shi.NJP.13.073016.2011} to discuss the quantum discord.
Then we arrive at the following conclusion.

\begin{proposition}
  Given a two-qubit state $\rho^{AB}$, assume that the matrix $R$ and the reduced density matrix $\rho^A$ are all invertible.
  Let $\tilde{\rho}^{AB}=(2\rho^A)^{-1/2}\rho^{AB}(2\rho^A)^{-1/2}$.
  Then the state $\tilde{\rho}^{AB}$ is isomorphic to the channel $\Lambda$, which is expressed in Bloch representation as $L_\Lambda=\tilde{R}^TR_+^T$,
  where $\tilde{R}$ and $R_+$ are, respectively, the $R$-matrix of the $\tilde{\rho}^{AB}$ and of the Bell state $\frac{1}{\sqrt2}(\ket{00}+\ket{11})$.
  The channel $L_\Lambda$, in turn, determines the quantum steering ellipsoid $\frak{E}$, which is given by $\bi{y}^T L_\Lambda^{-T}\eta L_\Lambda^{-1} \bi{y}=0$, or equivalently, $\bi{y}^TR^{-1}\eta R^{-T}\bi{y}=0$.
  Moreover, the ellipsoid $\frak{E}$ is invariant under any local filtering operation on qubit $A$.
\end{proposition}

We add a remark here.
In the above discussion, the channel $\Lambda$ (or $L_\Lambda$), which is isomorphic to the state $\tilde{\rho}^{AB}$, is applied on qubit $B$.
If we are concerned with qubit $A$, the quantum steering ellipsoid will be determined by the matrix $R^{-T}\eta R^{-1}$, just as we presented in \cite{Shi.NJP.13.073016.2011}.


\section{Quantum channel and concurrence of complement state} \label{sec: channel and concurrence}

In this section, we will show that the channel $L_\Lambda$ is closely related to the concurrence of the complement state of $\rho^{AB}$.
According to \cite{Koashi.PhysRevA.69.022309.2004},
the state $\rho^{BC}$ is complement to $\rho^{AB}$ when there exists a
tripartite pure state $\rho^{ABC}$ such that $\Tr_A(\rho^{ABC})=\rho^{BC}$ and
$\Tr_C(\rho^{ABC})=\rho^{AB}$.

Assume that the rank of $\rho^{AB}$ is four.
The eigendecomposition of $\rho^{AB}$ is
\[
  \rho^{AB}=\sum_{i=0}^{3}\lambda_i\ket{\psi_i}\bra{\psi_i}.
\]
We need a four-level quantum system, denoted by qudit $C$, to purify $\rho^{AB}$.
The purification of $\rho^{AB}$ is
\begin{equation} \label{PsiABC form AB and C}
  \ket{\Psi}=\sum_{i=0}^{3}\sqrt{\lambda_i}\ket{\psi_i}\otimes\ket{i^C},
\end{equation}
where $\ket{i^C}$ for $i=0,1,2,3$ constitute the basis of the Hilbert space of qudit $C$.
Then $\rho^{BC}=\Tr_A\ket{\Psi}\bra{\Psi}$ is a $2\otimes4$ rank-2 state and is the complement to $\rho^{AB}$.

Note that (\ref{PsiABC form AB and C}) can be regarded as the Schmidt decomposition with respect to the bipartite partition $AB:C$.
We can also perform Schmidt decomposition with respect to the partition $A:BC$.
Assume that the eigendecomposition of $\rho^{BC}$ is
\begin{equation}
  \rho^{BC}=\alpha_0\ket{\varphi_0}\bra{\varphi_0}+\alpha_1\ket{\varphi_1}\bra{\varphi_1}.
\end{equation}
It follows that
\begin{equation} \label{PsiABC form A and BC}
  \ket{\Psi}=\sqrt{\alpha_0}\,\ket{\alpha_0}\otimes\ket{\varphi_0}+
    \sqrt{\alpha_1}\,\ket{\alpha_1}\otimes\ket{\varphi_1},
\end{equation}
where $\ket{\alpha_i}$ for $i=0,1$ are the eigenvectors of $\rho^A$.
In fact $\alpha_i$ are the eigenvalues of $\rho^A$ associated with $\ket{\alpha_i}$.

To proceed to evaluate the concurrence of $\rho^{BC}$, we need the following two definitions, which are from \cite{Hildebrand.JMP.48.102108.2007}.

\begin{definition}
  For any $n\times n$ matrix $A$, the second symmetric function $S_2(A)$ is defined as
  \begin{equation}
    S_2(A)=\frac{1}{2}[(\Tr A)^2-\Tr A^2].
  \end{equation}
\end{definition}

Let $H^d$ be the space of complex Hermitian matrices of size $d\times d$ and let $H_+^d$ be the cone of positive semidefinite matrices.
The concurrence of a bipartite state is given by
\begin{definition}
  Given a $d_1\otimes d_2$ bipartite state $\rho$, denote by $\rho_1$ and $\rho_2$ the reduced state of the first and the second subsystem, respectively, i.e., $\rho_{1(2)}=\Tr_{2(1)}\rho$.
  The concurrence $\mathrm{Con}(\rho)$ for the state $\rho$ is defined on $H_+^{d_1d_2}$ as the convex roof of the function $2\sqrt{S_2(\rho_1)}$ or $2\sqrt{S_2(\rho_2)}$.
\end{definition}

Now let us consider the concurrence of the $2\otimes4$ bipartite state $\rho^{BC}$.
The first step is to write out the reduced state $\rho^B$.
Let us consider a more general case: an arbitrary Hermitian operator $X^{BC}$
on the two-dimension Hilbert space $\mathcal{H}_2$ spanned by the two eigenvectors of $\rho^{BC}$, i.e., $\mathcal{H}_2=\frak{span}\{\ket{\varphi_0},\ket{\varphi_1}\}$. The reduced operator $X^B=\Tr_CX^{BC}$.
Define the following four operators on $\mathcal{H}_2$.
\numparts
\begin{eqnarray}
  \varsigma_0=\ket{\varphi_0}\bra{\varphi_0}+\ket{\varphi_1}\bra{\varphi_1}, \quad \varsigma_1=\ket{\varphi_0}\bra{\varphi_1}+\ket{\varphi_1}\bra{\varphi_0},
  \label{generalized Pauli operators 12}\\
  \varsigma_2=-i\ket{\varphi_0}\bra{\varphi_0}+i\ket{\varphi_1}\bra{\varphi_0}, \quad \varsigma_3=\ket{\varphi_0}\bra{\varphi_0}-\ket{\varphi_1}\bra{\varphi_1}.
  \label{generalized Pauli operators 34}
\end{eqnarray}
\endnumparts
These four operators are analogous to the identity matrix $\sigma_0$ and the Pauli matrices $\sigma_i$ with $i=1,2,3$.
We can call them generalized Pauli operators.
Then we can express $X^{BC}$ as
\begin{equation}
  X^{BC}=\frac{1}{2}\sum_{\mu=0}^{3}x_\mu\varsigma_\mu,
\end{equation}
with $x_\mu$ real numbers.

To express $X^B$ explicitly, we have to calculate $\Tr_C\ket{\varphi_\mu}\bra{\varphi_\nu}$ for $\mu,\nu=0,\cdots,3$.
For this purpose, we resort to a maximally entangled state of the bipartite system $A:BC$, that is,
\begin{equation}
  \ket{\tilde{\Psi}}=\frac{1}{\sqrt2}(\ket{0^A}\otimes\ket{\varphi_0}
    +\ket{1^A}\otimes\ket{\varphi_1}).
\end{equation}
The state $\ket{\tilde{\Psi}}$ can be obtained from $\ket{\Psi}$ given by \eref{PsiABC form A and BC} by applying local filtering on qubit $A$, followed by an unitary operation:
\[
  \ket{\tilde{\Psi}}=[(UF)\otimes\mathbbm1_4]\ket{\Psi},
\]
where $F=\frac{1}{\sqrt{2\alpha_0}}\ket{\alpha_0}\bra{\alpha_0}
+\frac{1}{\sqrt{2\alpha_1}}\ket{\alpha_1}\bra{\alpha_1}$, and the unitary $U$ transforms $\ket{\alpha_0}$ and $\ket{\alpha_1}$ to $\ket{0^A}$ and $\ket{1^A}$, respectively.
Now we can see that
\[
  \ket{\varphi_i}\bra{\varphi_j}
    =2\bra{i^A}\big[\ket{\tilde{\Psi}}\bra{\tilde{\Psi}}\big]\ket{j^A},
\]
for $i,j=0,1$.
It follows that
\[
  \Tr_C\ket{\varphi_i}\bra{\varphi_j}
    =2\bra{i^A}\big[\Tr_C\ket{\tilde{\Psi}}\bra{\tilde{\Psi}}\big]\ket{j^A}
    =2\bra{i^A}\tilde{\rho}^{AB}\ket{j^A},
\]
with $\tilde{\rho}^{AB}=\Tr_C\ket{\tilde{\Psi}}\bra{\tilde{\Psi}}$.
Considering the generalized Pauli operators given by
\eref{generalized Pauli operators 12} \eref{generalized Pauli operators 34},
we have
\begin{eqnarray*}
  \Tr_C\varsigma_0=2\Tr_A[(\sigma_0\otimes\mathbbm1_2)\tilde{\rho}^{AB}], \quad
  \Tr_C\varsigma_1=2\Tr_A[(\sigma_1\otimes\mathbbm1_2)\tilde{\rho}^{AB}], \\
  \Tr_C\varsigma_2=-2\Tr_A[(\sigma_2\otimes\mathbbm1_2)\tilde{\rho}^{AB}], \quad
  \Tr_C\varsigma_3=2\Tr_A[(\sigma_3\otimes\mathbbm1_2)\tilde{\rho}^{AB}].
\end{eqnarray*}
Now we can express $X^B$ as
\begin{eqnarray}
  X^B & =\Tr_CX^{BC}=\frac{1}{2}\sum_{\mu=0}^{3}x_{\mu}\Tr_C\varsigma_{\mu} \nonumber \\
    & =\Tr_A\big\{[(x_0\sigma_0+x_1\sigma_1-x_2\sigma_2+x_3\sigma_3)
      \otimes\mathbbm1]\,\tilde{\rho}^{AB}\big\}.
\end{eqnarray}
Straightforward calculation shows that
\begin{equation} \label{XB}
  X^B=\frac{1}{2}(\bi{x}^T R_+\tilde{R})\cdot\bm{\sigma}
    =\frac{1}{2}(\bi{x}^T L_\Lambda^T)\cdot\bm{\sigma},
\end{equation}
where $\bm{\sigma}=(\sigma_0,\sigma_1,\sigma_2,\sigma_3)$, and in the last equation we have used \eref{L Lambda}.

In order to obtain the concurrence of $\rho^{BC}$, we use the following theorem, which is essentially from \cite{Hildebrand.JMP.48.102108.2007}.
\begin{theorem} \label{theorem from Hildebrand}
  Let $X$ be a $d_1\otimes d_2$ bipartite matrix of rank not exceeding $2$.
  Further, let $\mathcal{H}_2\subset\mathbb{C}^{d_1d_2}$ be a linear complex subspace of dimension $2$ such that the range of $X$ is contained in $\mathcal{H}_2$.
  Denote by $H^{d_1d_2}$ the space of all Hermitian matrices on $\mathbb{C}^{d_1d_2}$.
  Let $H_2$ be the subspace of all matrices in $H^{d_1d_2}$ whose range is contained in $\mathcal{H}_2$.
  Define the two quadratic forms $Q_1:A\mapsto 2[(\Tr A)^2-\Tr(\Tr_2A)^2]$ and
  $Q_2:A\mapsto (\Tr A)^2-\Tr A^2$ on $H^{d_1d_2}$.

  Then the generalized eigenvalues of the pencil $Q_1|_{H_2}-wQ_2|_{H_2}$ are all real.
  Denote them by $w_1$, $w_2$, $w_3$ and $w_4$ in decreasing order.
  Then the concurrence of $X$ is given by
  \begin{equation}
    \mathrm{Con}(X)=\sqrt{Q_1(X)-w_2Q_2(X)}.
  \end{equation}
\end{theorem}

In the case we are considering, $Q_1(X)$ is given by
\begin{eqnarray*}
  Q_1(X^{BC}) & =2[(\Tr X^{BC})^2-\Tr(\Tr_C X^{BC})^2] \\
    & =2(\Tr X^B)^2-2\Tr(X^B)^2 \\
    & =2x_0^2-(\bi{x}^T L_\Lambda^T)^2 \\
    & =\bi{x}^TL_\Lambda^T\eta L_\Lambda \bi{x},
\end{eqnarray*}
where in the third equation we have used \eref{XB}.
Then the matrix form of $Q_1$ is
\begin{equation}
  Q_1=L_\Lambda^T\eta L_\Lambda.
\end{equation}
As for $Q_2(X)$, we have $Q_2(X^{BC})=\frac{1}{2}\bi{x}^T\eta\bi{x}$.
It follows that the matrix form of $Q_2$ is given by $Q_2=\frac{1}{2}\eta$.

According to Theorem \ref{theorem from Hildebrand}, we have to solve the following equation to find $w_2$.
\begin{equation}
  \det\Big(L_\Lambda^T\eta L_\Lambda-\frac{w}{2}\eta\Big)=0.
\end{equation}
It is equivalent to
\begin{equation}
  \det\Big(\tilde{R}\eta \tilde{R}^T-\frac{w}{2}\eta\Big)=0.
\end{equation}
Or in other words, $w$ is the eigenvalue of $2\tilde{R}\eta\tilde{R}^T\eta$.

Obviously, the channel $L_\Lambda$ plays the central role in the evaluation of $\mathrm{Con}(\rho^{BC})$.
It should be mentioned that one can employ the approach presented in \cite{Hellmund.PhysRevA.79.052319.2009} to get the same results.

\section{Purification and Koashi-Winter relation} \label{sec: KW relation}
Given a bipartite state $\rho^{AB}$, by defining its complement state $\rho^{BC}$, K-W relation \cite{Koashi.PhysRevA.69.022309.2004} establishes the connection between the classical correlation of the former (i.e., $\rho^{AB}$) and the EoF of the latter (i.e., $\rho^{BC}$).
This relation gives new insight into both of the optimal ensemble with respect to the classical correlation or the quantum discord and that with respect to the EoF.
We outline briefly the K-W relation and make some remarks as follows.

We restrict our attention to the case of two-qubit states.
Given a $\rho^{AB}$ with rank $r$ ($r=2,3,4$), we add a $r$-dimensional ancilla $C$.
The purification of $\rho^{AB}$ is then given by
\begin{equation}
  \ket{\Psi}=\sum_{i=0}^{r-1}\sqrt{q_i}\otimes\ket{\psi_i^{AB}}\ket{i^C},
\end{equation}
where the probabilities $q_i$ and the pure states $\ket{\psi_i^{AB}}$ constitute an ensemble for $\rho^{AB}$,
and $\{\ket{i^C}\}_{i=0,\cdots,r-1}$ is the basis of the Hilbert space of particle $C$.
Note that $\ket{\psi_i^{AB}}$ may not be the eigenstates of $\rho^{AB}$ and may not be orthogonal to one another, and that the rank-two reduced state $\rho^{BC}=\Tr_A\ket{\Psi}\bra{\Psi}$ is the complement state of $\rho^{AB}$.

\begin{figure}[bpth]
\begin{center}
  \includegraphics[width=0.7\textwidth]{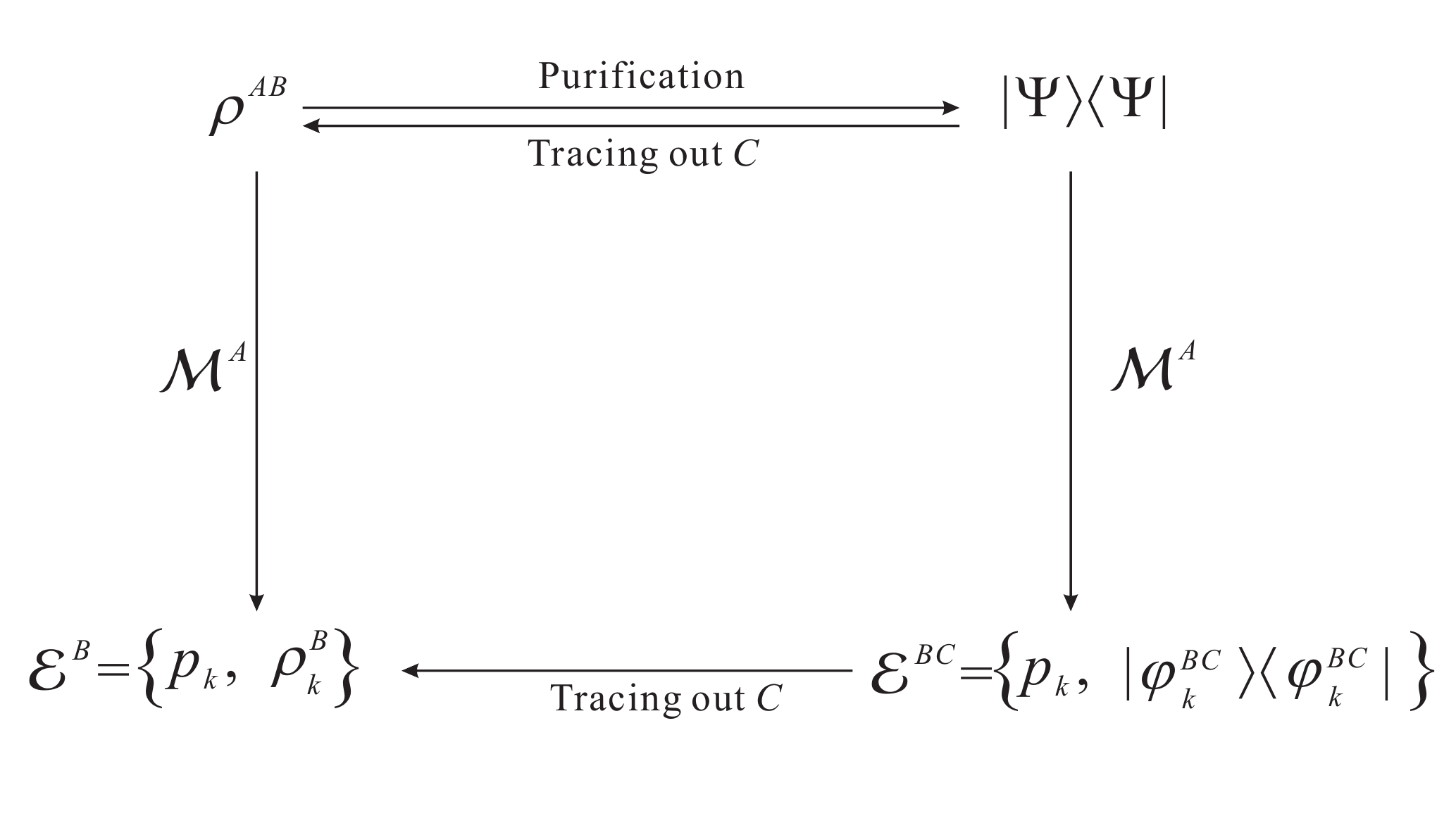}
\end{center}
\caption{The 2-qubit state $\rho^{AB}$ is purified into a tripartite pure state $\ket{\Psi}$. For $\rho^{AB}$, performing a POVM measurement $\mathcal{M}^A$ on qubit $A$ induces an ensemble for qubit $B$, namely $\mathscr{E}^B$. For $\ket{\Psi}$, the $\mathcal{M}^B$ induces the ensemble for the composite system $BC$, namely $\mathscr{E}^{BC}$. Tracing out particle $C$ from each member in $\mathscr{E}^{BC}$ will give rise to $\mathscr{E}^{B}$.}
  \label{fig: KW relation}
\end{figure}

Suppose that an $n$-element POVM measurement $\mathcal{M}=\{M_k\}$ with ${k=0,\cdots,n-1}$ is performed on qubit $A$.
We need only consider the case that all $M_k$ are of rank one \cite{D'Ariano.JPhysA.38.5979.2005,Hamieh.PhysRevA.70.052325.2004}, that is, $M_k$ is proportional to the one-dimensional projector.
From the viewpoint of the whole system in the pure state $\ket{\Psi}$, the measurement $\mathcal{M}$ gives rise to an ensemble for $\rho^{BC}$, denoted by $\mathscr{E}^{BC}=\{p_k,\ket{\varphi_k^{BC}}\}$, where
\begin{eqnarray*}
  & p_k=\bra{\Psi}(M_k\otimes\mathbbm1_2\otimes\mathbbm1_r)\ket{\Psi}, \\
  & \ket{\varphi^{BC}_k}\bra{\varphi^{BC}_k}
    =\Tr_A\big[(M_k\otimes\mathbbm1_2\otimes\mathbbm1_r)\,\ket{\Psi}\bra{\Psi}\big]/p_k,
\end{eqnarray*}
with $\mathbbm1_r$ being $r\times r$ identity matrix.
From the viewpoint of the state $\rho^{AB}$, the measurement $\mathcal{M}$ on $A$ gives rise to an ensemble for $\rho^B$, that is, $\mathscr{E}^B=\{p_k,\rho_k^B\}$, where
\begin{eqnarray*}
  & p_k=\Tr\big[(M_k\otimes\mathbbm1_2)\rho^{AB}\big], \\
  & \rho_k^B=\Tr_A\big[(M_k\otimes\mathbbm1_2)\rho^{AB}\big]/p_k.
\end{eqnarray*}
It is easy to see that the ensemble $\mathscr{E}^B$ can be induced from $\mathscr{E}^{BC}$, namely, $\rho^{B}_k=\Tr_C\big(\ket{\varphi_k^{BC}}\bra{\varphi_k^{BC}}\big)$. In figure \ref{fig: KW relation}, we illustrate the relations just mentioned.

The EoF of each pure $\ket{\varphi^{BC}_k}$ is given by $E(\varphi^{BC}_k)=S(\rho^B_k)$.
The average EoF over the ensemble $\mathscr{E}^{BC}$ is given by
\[
  \overline{E}^{BC}=\sum_{k=0}^{n-1}p_kE(\varphi^{BC}_k)=\sum_{k=0}^{n-1}p_kS(\rho^B_k).
\]
It follows that the average entropy of the ensemble $\mathscr{E}^B$ is equal to the average EoF:
\begin{equation}\label{pre-KW relation}
  \overline{S}=\overline{E}^{BC}.
\end{equation}
Note that (\ref{pre-KW relation}) depends on the measurement on $A$.
The dependence is removed by by performing minimization over all POVM measurements.
It then follows that
$\overline{S}_{\min}=\overline{E}^{BC}_{\min}$.
Koashi and Winter pointed out that the $\overline{E}^{BC}_{\min}$ obtained in this way is exactly the EoF of $\rho^{BC}$.
Then the K-W relation is expressed as
\begin{equation} \label{min S eq eof}
    \overline{S}_{\min}=E(\rho^{BC}).
\end{equation}
It follows that
\begin{eqnarray}
  C=S(\rho^B)-E(\rho^{BC}), \label{C in EoF}\\
  D=S(\rho^A)+E(\rho^{BC})-S(\rho^{AB}). \label{D in EoF}
\end{eqnarray}

Let's make some remarks as follow.

(i) Conventionally the EoF is regarded as coming from mathematical consideration. Now we see from K-W relation that EoF has relevance to the measurements.

(ii) If $\mathrm{rank}(\rho^{AB})=2$, then $\rho^{BC}$ is a two-qubit rank-2 state.
It follows from \cite{Hill.PhysRevLett.78.5022.1997} that $E(\rho^{BC})$ is achieved on the 2-state ensemble $\mathscr{E}^{BC}$ such that
$E(\rho^{BC})=E(\varphi^{BC}_0)=E(\varphi^{BC}_1)$.
This means that we need only perform the two-element POVM measurement, namely, von Neumann measurement, on qubit $A$ in order to obtain $\overline{S}_{\min}$ and thereby the discord $D$.
Moreover, the optimal measurement will induce an equi-entropy decomposition of $\rho^B$, namely, $S(\rho_0^B)=S(\rho_1^B)$.

(iii) In the case of $\mathrm{rank}(\rho^{AB})=3$ or $4$, although the state $\rho^{BC}$ is of rank two, there is no definite answer as to whether a two-state ensemble will yield the EoF of $\rho^{BC}$.
To put it in terms of the discord $D$ (or the MAE $\overline{S}_{\min}$), there is no definite answer as to whether the optimal measurement performed on qubit $A$ is the von Neumann measurement.
Therefore, strictly speaking, the MAE obtained by means of von Neumann measurement only provides the upper bound of the EoF of $\rho^{AC}$.


\section{Two-qubit rank-two states} \label{sec: rank two states}

According to the above discussion, we see that the discord of any two-qubit rank-2 state can be evaluated exactly. In this section, we provide the analytic results and the geometric description.

\subsection{Analytic results} \label{subsec: exact results of R2 state}

The eigendecomposition of $\rho^{AB}$ is
\begin{equation} \label{rho AB}
    \rho^{AB}=\lambda_0\ket{\psi_0}\bra{\psi_0}+\lambda_1\ket{\psi_1}\bra{\psi_1}.
\end{equation}
From the conclusion in \cite{Walgate.PhysRevLett.85.4972.2000}, we know that the eigenstates $\ket{\psi_0}$ and $\ket{\psi_1}$ can be simultaneously transformed to the following forms by local unitary operations.
\begin{equation}\label{psii in rhoAB}
  \ket{\psi_0}=a_0\ket{0}\ket{0}+b_0\ket{\eta}\ket{1}, \quad
  \ket{\psi_1}=a_1\ket{1}\ket{0}+b_1\ket{\eta^\perp}\ket{1},
\end{equation}
where $|a_k|^2+|b_k|^2=1$ for $k=0,1$, and $\ket{\eta}$ is orthogonal to $\ket{\eta^\perp}$. Let
\begin{equation} \label{etas in rhoAB}
  \ket{\eta}=c\ket{0}+d\ket{1}, \quad
  \ket{\eta^\perp}=-d^*\ket{0}+c^*\ket{1},
\end{equation}
with $|c|^2+|d|^2=1$.

Attaching a qubit $C$ to the two qubits $A$ and $B$, we write the purification of $\rho^{AB}$ as
\begin{equation*}
    \ket{\Psi}=\sqrt{\lambda_0}\ket{\psi_0}\otimes\ket{0}
                +\sqrt{\lambda_1}\ket{\psi_1}\otimes\ket{1}.
\end{equation*}
The reduced state $\rho^{BC}$ is a two-qubit rank-2 state, and the concurrence can be obtained easily.
The square of the concurrence is given by
\begin{eqnarray*}
    & [\mathrm{Con}(\rho^{BC})]^2 \\
   =& 2\lambda_0\lambda_1\Big[|a_0b_1 c^*-a_1b_0c|^2
                 +2|d|^2\big(|a_0|^2|b_1|^2+|a_1|^2|b_0|^2\big)\Big] \\
    & -2\lambda_0\lambda_1\,
      \Big|(a_0b_1 c^*-a_1b_0c)^2-4a_0a_1b_0b_1|d|^2\Big|.
\end{eqnarray*}
The EoF of $\rho^{BC}$ is then
\begin{equation} \label{EoF of rhoBC}
    E(\rho^{BC})=H\bigg(\frac{1+\sqrt{1-[\mathrm{Con}(\rho^{BC})]^2}}{2}\,\bigg),
\end{equation}
where the function $H(x)$ is defined as
\begin{equation} \label{def: function h}
    H(x)=-x\log_2x-(1-x)\log_2(1-x),
\end{equation}
for $x\in[0,1]$.
From (\ref{C in EoF}) and (\ref{D in EoF}) we obtain the analytic expressions for the classical correlation $C$ and quantum discord $D$.

For later reference, we need the expression for $[\mathrm{Con}(\rho^{BC})]^2$ in the case that the parameters $a_k$, $b_k$, $c$ and $d$ are assumed to be real numbers.
In this case, we have the following results.

(i) If $c^2(a_0b_1-a_1b_0)^2\geqslant 4a_0a_1b_0b_1d^2$, then
\begin{equation} \label{sq con1 rhoBC}
  [\mathrm{Con}_{\mathrm{(i)}}(\rho^{BC})]^2=4\lambda_0\lambda_1d^2(a_0b_1+a_1b_0)^2.
\end{equation}

(ii) If $c^2(a_0b_1-a_1b_0)^2< 4a_0a_1b_0b_1d^2$, then
\begin{equation} \label{sq con2 rhoBC}
  [\mathrm{Con}_{\mathrm{(ii)}}(\rho^{BC})]^2=4\lambda_0\lambda_1(a_0b_1-a_1b_0)^2.
\end{equation}
From (\ref{EoF of rhoBC}) we have the corresponding $E_{\mathrm{(i)}}(\rho^{BC})$ and $E_{\mathrm{(ii)}}(\rho^{BC})$. From (\ref{C in EoF}) and (\ref{D in EoF}), we have $C_{\mathrm{(i)}}$, $C_{\mathrm{(ii)}}$ and $D_{\mathrm{(i)}}$, $D_{\mathrm{(ii)}}$.

\subsection{Geometric picture for entangled states}

As stated in Section \ref{sec: KW relation}, for any two-qubit rank-2 state, the optimal measurement on qubit $A$ will realize an equi-entropy decomposition of the reduced state of qubit $B$.
We will illustrate this result in the geometric picture presented in \cite{Shi.NJP.13.073016.2011}.
In this subsection we discuss the case of entangled states.

For the entangled state $\rho^{AB}$, the $R$-matrix of $\rho^{AB}$ is invertible and thus the quantum steering ellipsoid $\frak{E}$ exists.
Its equation has been given in Section \ref{subsec: steering ellipsoid}, where the consideration is based on the notation of quantum channel.
As a comparison, let us spend a moment to give a brief outline of the original discussion presented in \cite{Verstraete.diss.2002}.

Suppose that a POVM measurement is performed on qubit $A$.
The postmeasurement state of qubit $B$, corresponding to the operation element $M_k$, is $\rho^B_k$.
In Hilbert-Schmidt space, we express $M_k$, $\rho_k^A$ as
\[
  M_k=\sum_{\mu=0}^{3}x_{k,\mu}\sigma_\mu, \quad
    \rho_k^B=\frac{1}{2}\sum_{\mu=0}^{3}y_{k,\mu}\sigma_\mu,
\]
where
\begin{equation*}
    x_{k,\mu}=\frac{1}{2}\Tr(M_k\,\sigma_\mu), \quad
    y_{k,\mu}=\Tr(\rho^B_k\,\sigma_\mu).
\end{equation*}
We define two 4-component vectors (in column form) $\bi{x}_k$ and $\bi{y}_k$ as
\begin{eqnarray*}
  \bi{x}_k=(x_{k,0},\,x_{k,1},\,x_{k,2},\,x_{k,3})^T, \label{4-vector x} \\
  \bi{y}_k=(y_{k,0},\,y_{k,1},\,y_{k,2},\,y_{k,3})^T. \label{4-vector y}
\end{eqnarray*}
Note that $y_{k,0}=1$ for all $k$ and the 3-component vector $\vec{y}_k=(y_{k,1},\,y_{k,2},\,y_{k,3})^T$ is just the Bloch vector of $\rho_k^B$.
Direct calculation shows the following equations.
\begin{equation}\label{relation between y and x}
  p_k\,\bi{y}_k=R^T\bi{x}_{k}, \quad
  p_k=\sum_{\mu=0}^{3}R_{\mu 0}\,x_{k,\mu},
\end{equation}
(\ref{relation between y and x}) provides the
relationship between the measurement on $A$ and the induced component in postmeasurement ensemble for the reduced state of $B$.
Recall that $R$ is invertible.
So the vector $\bi{y}_k$ is in a one-to-one correspondence to the vector $\bi{x}_{k}$.
It follows that $\bi{x}_k=p_kR^{-T}\bi{y}_k$.
The non-negativity of $M_k$ leads us to
\begin{equation}\label{ellipsoid}
  \bi{y}_k^{T}\,\big[R^{-1}\,\eta\,R^{-T}\big]\,\bi{y}_k\geqslant 0,
\end{equation}
which is just the ellipsoid equation given by \eref{ellipsoid in tilde R} or \eref{ellipsoid E}.

The steering ellipsoid yields concrete geometric picture when we minimize the
average entropy $\overline{S}$.
Denote by $\vec{r}^B=(r_1^B,\,r_2^B,\,r_3^B)^T$ the Bloch vector of $\rho^B$.
The state $\rho^B$ can be represented by a point $B$ with the coordinate given by $(r_1^B,\,r_2^B,\,r_3^B)$.
It can be seen that point $B$ is in the interior of $\frak{E}$.
If qubit $A$ is measured by a $n$-element POVM measurement, the $n$ Bloch vectors $\vec{r}_k^B$ ($k=0,\cdots,n-1$) of the postmeasurement states $\rho_k^B$ constitute a convex polytope $\mathcal{P}$, which is contained in the ellipsoid $\frak{E}$, that is, $B\in\mathcal{P}\subset\frak{E}$.
The average entropy is given by
\begin{equation*}
    \overline{S}=\sum_kp_kS(\rho_k^B)
      =\sum_kp_kH\Big(\frac{1+r_k^A}{2}\Big),
\end{equation*}
with $r_k^A=|\bi{r}^A_k|$ and the function $H$ given by (\ref{def: function h}).

By noting that both $\mathcal{P}$ and $\frak{E}$ are convex and the entropy function is concave, we see that the minimal value of $\overline{S}$ must be attained on the surface of $\mathfrak{E}$.
The surface of $\frak{E}$ represents all postmeasurement states of qubit $B$ when qubit $A$ is measured by rank-one POVMs.
Denoting by $\partial\frak{E}$ the surface of $\frak{E}$, we can write
\begin{equation}\label{min SB}
    \overline{S}_{\min}=\min_{\mathcal{M}}\sum_kp_kS(\rho^B_k)
    =\min_{\{p_k,\vec{r}_k^B\}}\sum_k p_k H\Big(\frac{1+r_k^B}{2}\Big),
\end{equation}
where the minimization is taken over all $\{p_k,\vec{r}_k^B\}$ with $\vec{r}_k^B\in\partial\frak{E}$ and $\sum_kp_k\vec{r}_k^B=\vec{r}^B$.

Now we proceed to give the geometric picture.
With $\rho^{AB}$ given by (\ref{rho AB})\ --\ (\ref{etas in rhoAB}), we let all parameters $a_k$, $b_k$, $c$ and $d$ take real numbers in order to simplify our calculations.
In this case the classical correlation and quantum discord have been obtained in Section \ref{subsec: exact results of R2 state}.
We present below the geometric description of the optimal decomposition of $\rho^B$.

Firstly let's define the following functions for later convenience.
\begin{eqnarray*}
     f_1=\lambda_0 a_0b_0+\lambda_1 a_1b_1, \quad
     f_2=\lambda_0 a_0b_0-\lambda_1 a_1b_1, \\
     f_3=\sqrt{\lambda_0\lambda_1}(a_1b_0+a_0b_1), \quad f_4=\sqrt{\lambda_0\lambda_1}(a_1b_0-a_0b_1), \\
     f_5=\sqrt{\lambda_0\lambda_1}(a_0a_1+b_0b_1), \quad f_6=\sqrt{\lambda_0\lambda_1}(a_0a_1-b_0b_1).
\end{eqnarray*}
From (\ref{ellipsoid}) the equation of steering ellipsoid $\mathfrak{E}$ is given by
\begin{eqnarray}
   &(c^2f_3^2+d^2f_1^2)y_1^2+(f_6^2+d^2f_1^2)y_3^2
     +2c f_3f_6y_1y_3 \nonumber\\
   & +(c^2f_4^2+d^2f_2^2)y_2^2-2c f_3f_5y_1-2f_5f_6y_3
     +(f_5^2-d^2f_1^2)=0 \label{ellipsoid 1}
\end{eqnarray}
In order to describe $\mathfrak{E}$ more clearly, we apply a rotation about $y_2$ axis:
\begin{equation}
    y_1\rightarrow y_1\cos\eta-y_3\sin\eta, \quad
    y_3\rightarrow y_1\sin\eta+y_3\cos\eta \label{rotation},
\end{equation}
with the rotation angle $\eta$ given by
\begin{equation}\label{angle eta}
    \sin\eta=-\frac{c f_3}{\sqrt{c^2f_3^2+f_6^2}}, \quad
    \cos\eta=\frac{f_6}{\sqrt{c^2f_3^2+f_6^2}}.
\end{equation}
Denote by $\mathfrak{E}'$ the ellipsoid after the rotation.
The equation of $\mathfrak{E}'$ is
\begin{eqnarray}
    & d^2f_1^2y_1^2+(c^2 f_4^2+d^2 f_2^2)y_2^2 \nonumber \\
    & \quad  +(f_6^2+c^2f_3^2+d^2f_1^2)
       \left[y_3-\frac{f_5\sqrt{f_6^2+c^2f_3^2}}
                      {f_6^2+c^2f_3^2+d^2f_1^2}\right]^2 \nonumber\\[8pt]
   =&\frac{d^2f_1^2\,(f_6^2-f_5^2+c^2f_3^2+d^2f_1^2)}
            {f_6^2+c^2f_3^2+d^2f_1^2}.
\end{eqnarray}
It can be seen that the $y_3$ axis is one of the symmetry axes of $\mathfrak{E}'$ and the center of $\mathfrak{E}'$ is located on the point
\begin{equation*}
    \bigg(0,\;0,\;\frac{f_5\sqrt{f_6^2+c^2f_3^2}}
                      {f_6^2+c^2f_3^2+d^2f_1^2}\bigg).
\end{equation*}

Before rotation, the Bloch vector of the local state $\rho^B$ is
\begin{equation*}
    \vec{r}^B=2c f_1\hat{e}_1+[\lambda_0(a_0^2-b_0^2)+\lambda_1(a_1^2-b_1^2)]\hat{e}_3,
\end{equation*}
where $\hat{e}_k$ denotes the unit vector along $y_k$ axis.
After the rotation, the Bloch vector $\vec{r}^B$ is transformed to
\begin{equation*}
    \vec{r}\,'^{B}=\left(
    \begin{array}{ccc}
    \cos\eta & 0 & \sin\eta \\
    0 & 1 & 0 \\
    -\sin\eta & 0 & \cos\eta
    \end{array}\right)
    \vec{r}^B,
\end{equation*}
with the the angle $\eta$ given by (\ref{angle eta}).
Obviously, $\vec{r}\,'^B$ is still in the $y_1y_3$ plane.

\begin{figure}[bhpt]
\begin{center}
    \includegraphics[width=0.6\textwidth]{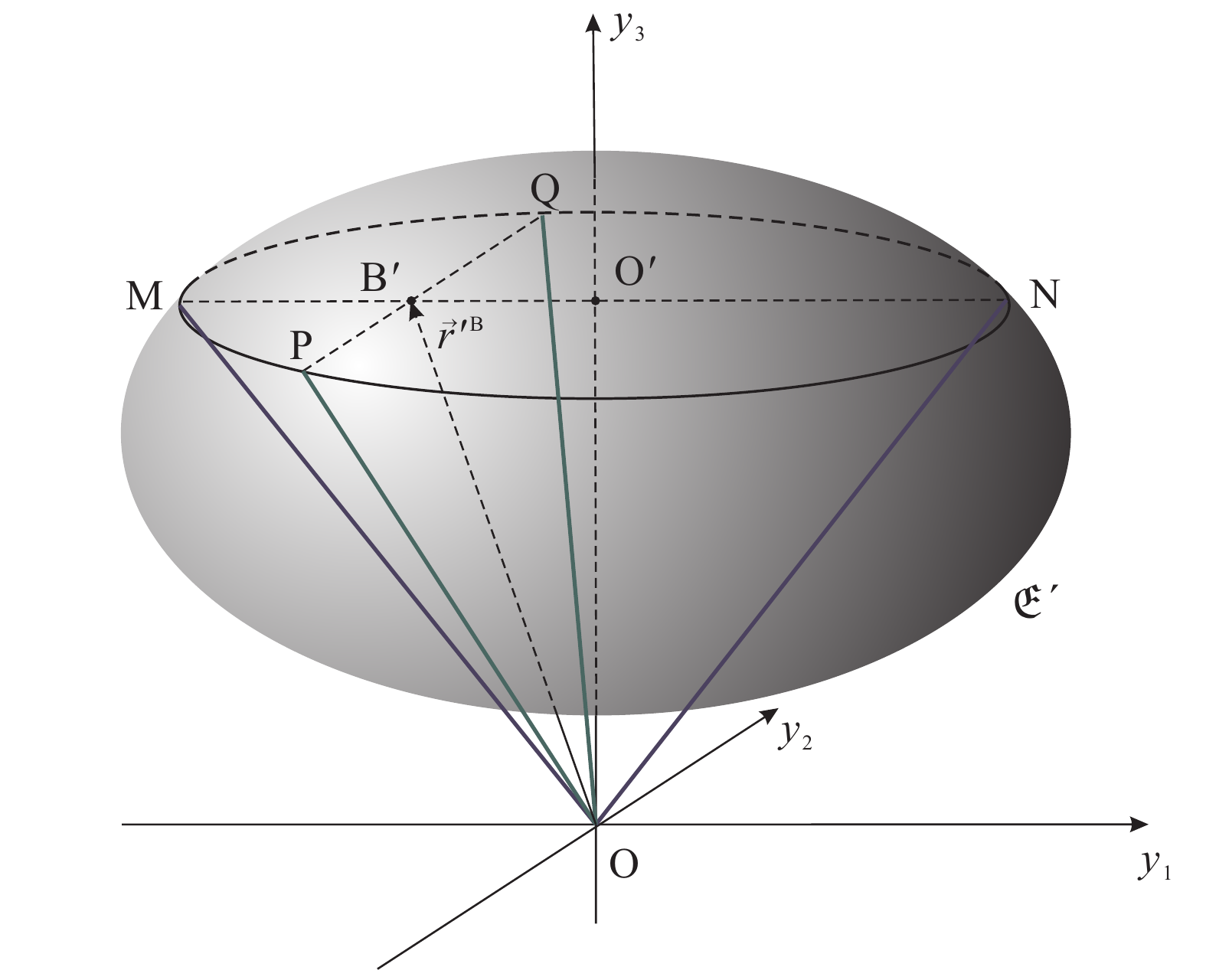}
  \caption{Schematic illustration of steering ellipsoid $\mathfrak{E}'$ for $\rho^{AB}$.
  Point $B'$ corresponds to vector $\vec{r}\,'^{B}$, which is the Bloch vector of $\rho^B$ after rotation given by (\ref{rotation}).
  A plane containing point $B'$ and parallel to $y_1y_2$ plane intersects the surface of $\mathfrak{E}'$ in a ellipse.
  Depending on the parameters of $\rho^{AB}$, the optimal ensemble corresponds to either of the following two decompositions of $\vec{r}\,'^B$: the convex combination of $\vec{r}_M$ and $\vec{r}_N$, or the convex combination of $\vec{r}_P$ and $\vec{r}_Q$.}
    \label{fig: ellipsoid}
\end{center}
\end{figure}

In figure \ref{fig: ellipsoid}, we illustrate the post-rotation ellipsoid $\frak{E}'$ and the post-rotation Bloch vector $\vec{r}\,'^B$.
Generally, the ellipsoid $\frak{E}'$ may not possess the rotational symmetry about any principle axis, and the Bloch vector $\vec{r}\,'^B$ may not lie along the $y_3$ axis.
Nonetheless, there exist two forms of equi-entropy decomposition of $\vec{r}\,'^B$, which we describe as follows.

Suppose a plane is parallel to $y_1y_2$ plane and passes through point $B'$.
This plane intersects the surface of $\frak{E}'$, and the the intersection is an ellipse.
Two chords of the ellipse, $MB'N$ and $PB'Q$, pass through point $B'$.
The former is parallel to $y_1$ axis and the latter parallel to $y_2$ axis.
We see that $OM=ON$ and $OP=OQ$.
Then the equi-entropy decompositions are given by
\begin{eqnarray}
  \vec{r}\,'^B=p_M\,\vec{r}_M+p_N\,\vec{r}_N, \label{decomposition MN}\\
  \vec{r}\,'^B=p_P\,\vec{r}_P+p_Q\,\vec{r}_Q, \label{decomposition PQ}
\end{eqnarray}
where $\vec{r}_M$ denotes the vector from point $O$ to $M$, and so forth.
It follows that the average entropy corresponding to the decomposition (\ref{decomposition MN}) is given by
\[
  \overline{S}_{MN}=p_M H\Big(\frac{1+r_M}{2}\Big)
    +p_N H\Big(\frac{1+r_N}{2}\Big)=H\Big(\frac{1+r_M}{2}\Big).
\]
Similarly, \eref{decomposition PQ} gives
\[
  \overline{S}_{PQ}=H\Big(\frac{1+r_P}{2}\Big).
\]

After tedious calculations, we get the length of $OM$ and $OP$:
\begin{eqnarray}
  & r_M^2=r_N^2=1-4\lambda_0\lambda_1d^2(a_0b_1+a_1b_1)^2, \label{OM sq eq ON sq}\\
  & r_P^2=r_Q^2=1-4\lambda_0\lambda_1(a_0b_1-a_1b_0)^2. \label{OP sq eq OQ sq}
\end{eqnarray}
Recalling (\ref{EoF of rhoBC}), (\ref{sq con1 rhoBC}) and (\ref{sq con2 rhoBC}), we have
\begin{equation} \label{MN PQ and EoF}
  \overline{S}_{MN}=E_{\mathrm{(i)}}(\rho^{BC}), \quad
  \overline{S}_{PQ}=E_{\mathrm{(ii)}}(\rho^{BC}).
\end{equation}
Hence we have shown that (\ref{decomposition MN}) or (\ref{decomposition PQ}), together with (\ref{OM sq eq ON sq}) and (\ref{OP sq eq OQ sq}), provide the geometric description of the optimal postmeasurement ensemble for $\rho^{B}$.

\subsection{Geometric picture for separable states}
If a two-qubit state $\rho^{AB}$ is separable, its steering ellipsoid may not exist. Nonetheless, the geometric description of the MAE is very clear, which we have presented in \cite{Shi.NJP.13.073016.2011}.
There we conjectured that equi-entropy decomposition is the necessary condition which must be satisfied in order that the average entropy reaches the minimum.
Now the discussion in Section \ref{sec: KW relation} shows that this conjecture is true.
A complete description is given below.

Any two-qubit separable state with rank two can be expressed, up to local unitary operations, as
\begin{equation}\label{state: rank two}
    \rho^{AB}=q\ket{0}\bra{0}\otimes\ket{0}\bra{0}
         +(1-q)\ket{\psi}\bra{\psi}\otimes\ket{\phi}\bra{\phi},
\end{equation}
where $q\in(0,1)$, $\ket{\psi}=\cos\alpha\ket{0}+\sin\alpha\ket{1}$,
and $\ket{\phi}=\cos\beta\ket{0}+\sin\beta\ket{1}$ with $\alpha,\beta\in(0,\frac{\pi}{2}]$. Here neither $\alpha$ or $\beta$ takes the value of zero, otherwise the state $\rho^{AB}$ is a trivial product state.
The purification of $\rho^{AB}$ is
\begin{equation}\label{state: purification of rho}
    \ket{\Psi}=\sqrt{q}\;\ket{0}\otimes\ket{0}\otimes\ket{0}
               +\sqrt{1-q}\;\ket{\psi}\otimes\ket{\phi}\otimes\ket{1}.
\end{equation}
The concurrence of $\rho^{BC}$ is easily calculated, that is,
$2\sqrt{q(1-p)}\cos\alpha\sin\beta$.
From K-W relation (\ref{min S eq eof}), we have
\begin{equation}\label{relation: Smin eq eof of rhoBC}
    \overline{S}_{\min}
      =H\bigg(\frac{1+\sqrt{1-4q(1-q)\cos^2\alpha\sin^2\beta}}{2}\;\bigg)
\end{equation}
Then quantum discord $D$ follows easily.

\begin{figure}[bhpt]
\begin{center}
\includegraphics[width=0.6\textwidth]{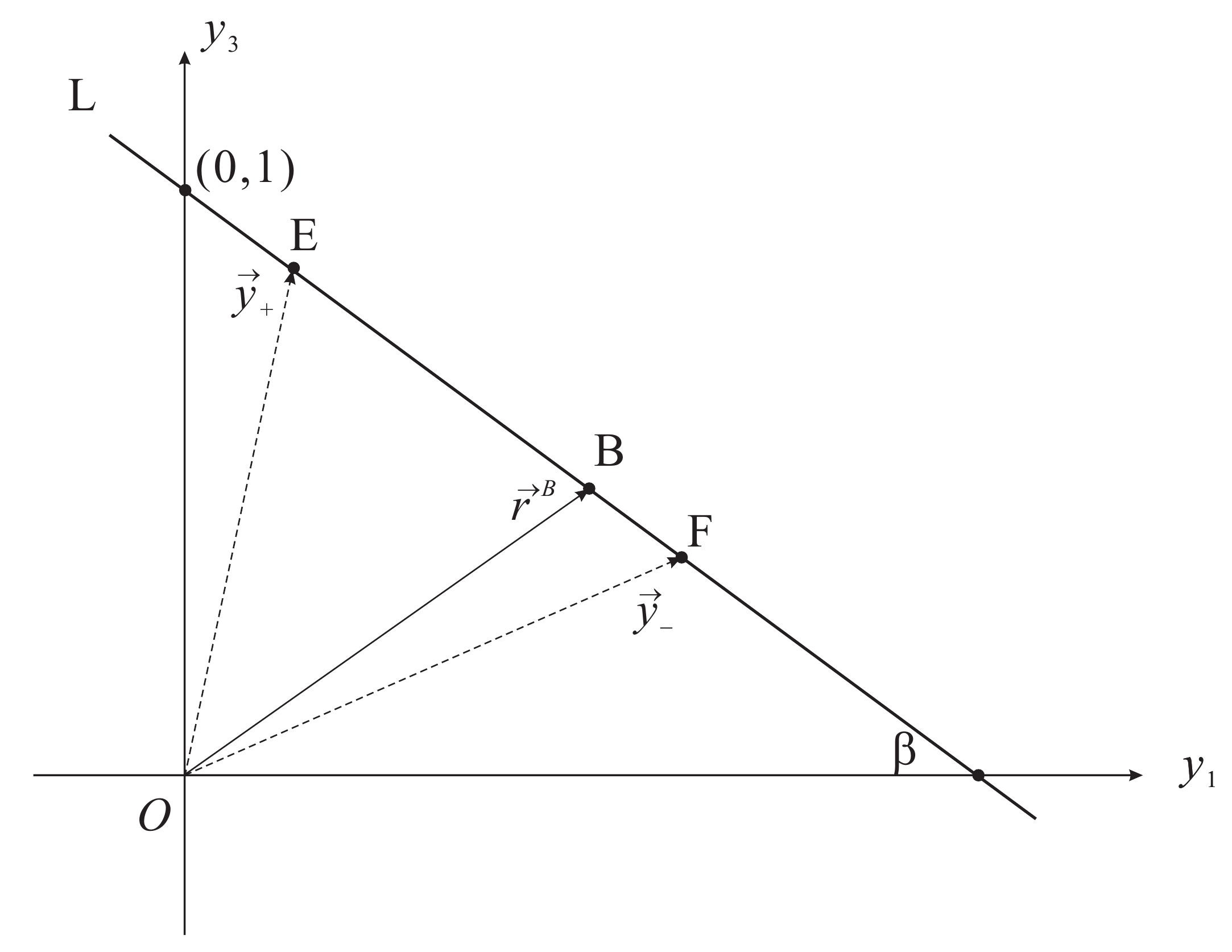}
 \caption{The postmeasurement states of qubit $B$, as well as the local state $\rho^B$ itself, is constrained on the line $L$, the equation of which is given by (\ref{line}).
 The optimal ensemble for qubit $B$ corresponds to the case in which the length of $OE$ is equal to that of $OF$ and further takes the maximal value.}\label{fig: line}
\end{center}
\end{figure}

Now we present the geometric description of the MAE.
Suppose that we perform von Neumann measurement on qubit $A$.
The measurement operators are given by
\begin{equation}
  M_\pm=\frac{\mathbbm{1}}{2}\pm x_1\sigma_x\pm x_2\sigma_y\pm x_3\sigma_z,
\end{equation}
with $x_1^2+x_2^2+x_3^2=1/4$.
Inserting the 4-component vectors
$\bi{x}_{\pm}=\big(1/2,\,\pm x_{1},\,\pm x_{2},\,\pm x_{3}\big)^T$
into (\ref{relation between y and x}) will give $p_{\pm}$ and $\bi{y}_{\pm}$.
The concrete forms of $p_{\pm}$ and $\bi{y}_{\pm}$ are a bit lengthy and we do not write them out explicitly.
It can be shown that the Bloch vectors $\vec{y}_{\pm}$ satisfy
\begin{equation}\label{line}
  y_{\pm,3}+y_{\pm,1}\tan\beta=1.
\end{equation}
Then the postmeasurement states of qubit $B$ is constrained on the line $L$, given by (\ref{line}).
The reduced state $\rho^B$ is also on this line (see figure \ref{fig: line}), as the Bloch vector of $\rho^B$ is given by
\begin{equation*}
    \vec{r}^B=\Big((1-q)\sin2\beta,\;0,\;q+(1-q)\cos2\beta\Big)^T,
\end{equation*}
which satisfies $r_3^B+r_1^B\tan\beta=1$.

Now we consider the equi-entropy decomposition of $\rho^B$, which means that
$\vec{r}^B=p_{+}\,\vec{y}_{+}+p_{-}\,\vec{y}_{-}$ with $|\vec{y}_{+}|=|\vec{y}_{-}|$.

Let $x_1=\frac{1}{2}\sin\theta\sin\varphi$, $x_2=\frac{1}{2}\cos\varphi$ and $x_3=\frac{1}{2}\cos\theta\sin\varphi$ with $0\leqslant\theta<2\pi$ and $0\leqslant\varphi<\pi$.
It is not difficult to see that either of the following conditions give rise to
$|\vec{y}_{+}|=|\vec{y}_{-}|$.
\begin{eqnarray}
  \sin\theta\cos\alpha-\cos\theta\sin\alpha=0, \label{cond: OE eq OF 1}\\
  q^2\cos^2\theta-(1-q)^2\cos^2(\theta-2\beta)=2q-1,
      \label{cond: OE eq OF 2}
\end{eqnarray}
The condition (\ref{cond: OE eq OF 1}) gives us a trivial result, namely, $\vec{y}_+=\vec{y}_-=\vec{r}^B$.
So we focus our attention on condition (\ref{cond: OE eq OF 2}).
By lengthy but straightforward calculations, we find that under (\ref{cond: OE eq OF 2}), the maximal value of $|\vec{y}_+|$ or $|\vec{y}_-|$ is given by
\begin{equation} \label{max OE}
    y_{\max}=\sqrt{1-4q(1-q)\cos^2\alpha\sin^2\beta}.
\end{equation}
Under this decomposition, we obtain the average entropy
\begin{equation} \label{entropy: average over EF}
    \overline{S}_{EF}=H\Big(\frac{1+y_{\max}}{2}\Big).
\end{equation}
Combining (\ref{relation: Smin eq eof of rhoBC}), (\ref{max OE}) and (\ref{entropy: average over EF}), we see that $\overline{S}_{\min}=\overline{S}_{EF}$.
Thus we attribute a geometric interpretation to the analytical expression for $\overline{S}_{\min}$.

\section{Two-qubit states with rank larger than two} \label{sec: larger rank}

If $\mathrm{rank}(\rho^{AB})>2$, the complement state $\rho^{BC}$ is a $2\otimes r$ state with $r=3$ or $4$.
Although $\mathrm{rank}(\rho^{BC})=2$, we cannot ensure that $E(\rho^{BC})$ is obtained on a two-element ensemble.
In this situation, if we performed von Neumann measurement on $A$ and acquire the MAE about qubit $B$, the MAE obtained in this way would not be the true EoF of $\rho^{BC}$, but rather an upper bound for $E(\rho^{BC})$.

Despite the fact that we have no definite results about $E(\rho^{BC})$, the upper and the lower bound are known.
In \cite{Osborne.PhysRevA.72.022309.2005}, the $I$-tangle $\tau$, an entanglement measure proposed by Rungta \etal in \cite{Rungta.PhysRevA.64.042315.2001}, is calculated exactly for any bipartite states with rank two.
For the state $\rho^{BC}$ under consideration, the following inequality holds \cite{Osborne.PhysRevA.72.022309.2005}.
\begin{equation} \label{Osborne inequality}
  E(\rho^{BC})\leqslant H\Big(\frac{1}{2}+\frac{1}{2}\sqrt{1-\tau(\rho^{BC})}\,\Big).
\end{equation}
On the other hand, the exact concurrence of $\rho^{BC}$ is available from the approach presented in \cite{Hildebrand.JMP.48.102108.2007,Hellmund.PhysRevA.79.052319.2009} (see also the Theorem \ref{theorem from Hildebrand}).
The following inequality is from \cite{Hellmund.PhysRevA.79.052319.2009}.
\begin{equation} \label{Hellmund inequality}
  E(\rho^{BC})\geqslant H\Big(\frac{1}{2}
    +\frac{1}{2}\sqrt{1-\mathrm{Con}^2(\rho^{BC})}\,\Big).
\end{equation}
Recall the K-W relation $\overline{S}_{\min}=E(\rho^{BC})$.
The inequality \eref{Osborne inequality} and \eref{Hellmund inequality} give the upper and lower bound for the MAE respectively.
In \cite{Yu.arXiv.1102.1301}, the authors provide the upper and lower bound for the quantum discord from this consideration.

Below we discuss two issues.
One will concern the upper bounds for the EoF of the rank-two state $\rho^{BC}$,
and the other the more general measurements performed on qubit $A$ to achieve the MAE $\overline{S}_{\min}$.

\begin{figure}
\begin{center}
  \includegraphics[width=0.7\textwidth]{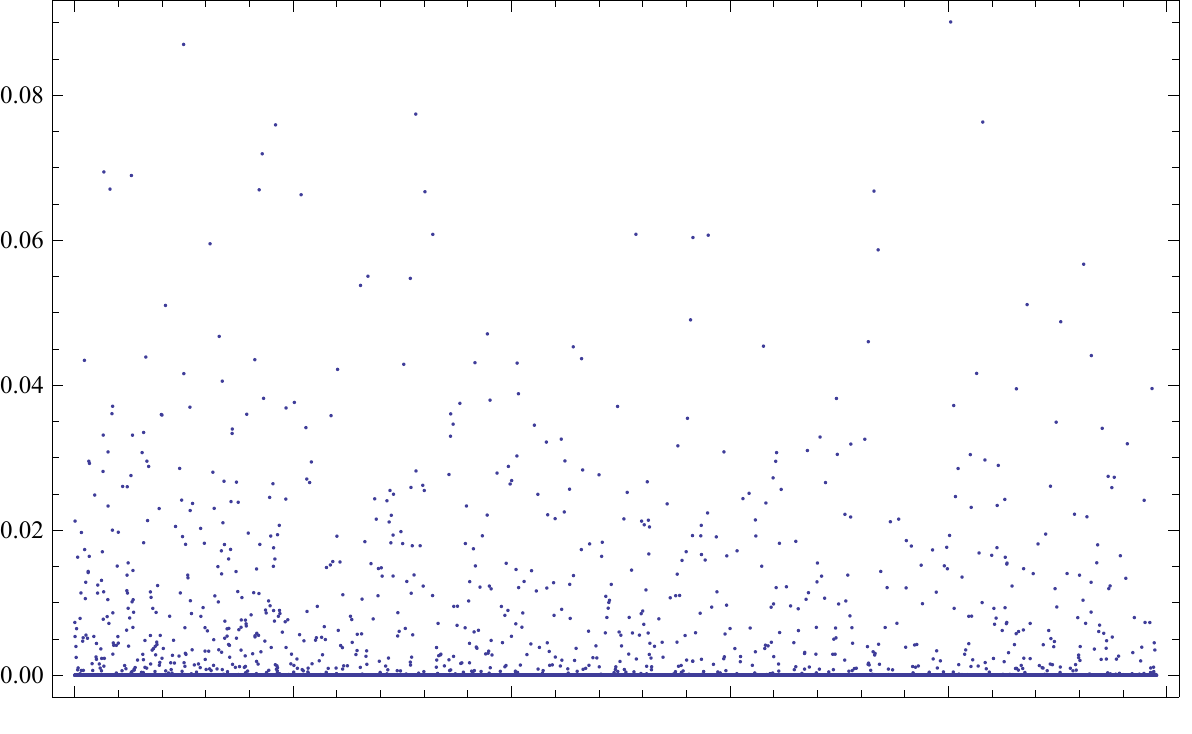}
  \caption{For about $10^{4}$ randomly selected X states $\rho^{AB}$ with rank three or four, we calculate the MAE $\overline{S}^{(2)}_{\min}$ for $\rho^{AB}$ and the $H(\tau)$ for the complement states $\rho^{BC}$.
  The difference $\Delta=H(\tau)-\overline{S}^{(2)}_{\min}$ is plotted in the figure.}\label{fig: AVEvsOs}
\end{center}
\end{figure}

Suppose that we are restricted to make von Neumann measurements on qubit $A$.
Let $\overline{S}^{(2)}_{\min}$ be the MAE of the postmeasurement ensemble for $\rho^B$.
It is an upper bound for $E(\rho^{BC})$, i.e., $E(\rho^{BC})\leqslant\overline{S}^{(2)}_{\min}$.
Note that the right-hand side of the inequality \eref{Osborne inequality}, denoted by $H(\tau)$ for short, is another upper bound for $E(\rho^{BC})$.
It is claimed that the expressions on the left-hand and right-hand side of (\ref{Osborne inequality}) usually differ only by about $10^{-4}$.
Let us compare these two upper bound.
We select randomly X states with rank three or four, calculate $\overline{S}^{(2)}_{\min}$ and $H(\tau)$, and plot the difference $\Delta=H(\tau)-\overline{S}^{(2)}_{\min}$ in figure \ref{fig: AVEvsOs}.
In calculating $\overline{S}^{(2)}_{\min}$, we adopt the approach provided in \cite{Ali.PhysRevA.81.042105.2010}.
That is, the optimization only concerns the equi-entropy decomposition and quasi-eigendecomposition of $\rho^{B}$ \cite{Shi.NJP.13.073016.2011}.
It should be mentioned that this approach is not strictly correct \cite{Lu.PhysRevA.83.012327.2011,Chen.arXiv.1102.0181,Girolami.PhysRevA.83.052108.2011}.

In figure \ref{fig: AVEvsOs} we see that $\Delta\geqslant0$, which means that $\overline{S}^{(2)}_{\min}$ is a tighter upper bound for $E(\rho^{BC})$.
For most of X states, $\Delta$ is about $10^{-15}$.
In fact the two upper bounds are equal to each other for these states.
However there are some state for which the difference $\Delta$ can be about $10^{-2}$, much larger than $10^{-4}$.
Further study shows that, (i) if $\overline{S}^{(2)}_{\min}$ is attained from the equi-entropy decomposition of $\rho^{B}$, then $\Delta=0$;
(ii) if $\overline{S}_{\min}$ is attained from the quasi-eigendecomposition, then the difference $\Delta$ is obviously larger than zero.

\begin{figure}
\begin{center}
  \includegraphics[width=0.7\textwidth]{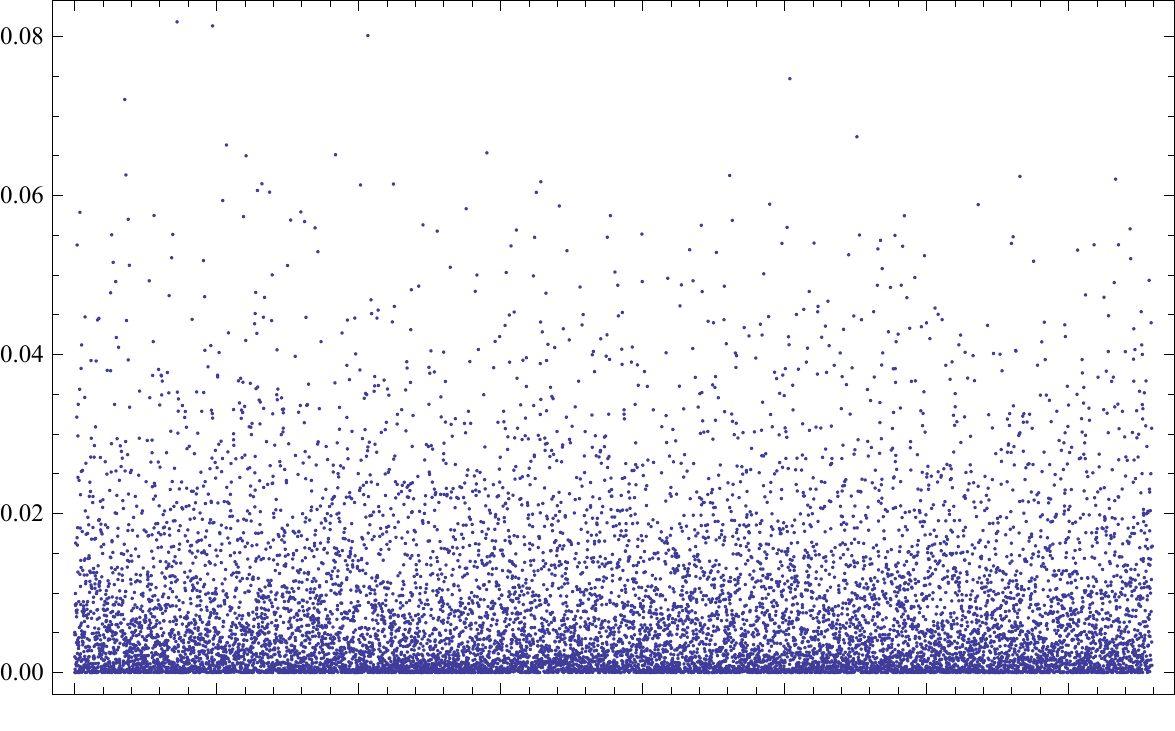}
  \caption{The numerical results of $\Delta$ for about $10^4$ randomly selected 2-qubit states.}\label{fig: general}
\end{center}
\end{figure}

For general two-qubit states, we calculate numerically the difference $\Delta$ for about $10^4$ randomly selected states and plot the results in figure \ref{fig: general}.
There are more states for which $\Delta>0$, meaning that the MAE is indeed a tighter upper bound for the EoF of the complement state.

Now let us turn to considering the general POVM measurements on qubit $A$.
In \cite{Hellmund.PhysRevA.79.052319.2009}, the authors studied the entanglement entropy for axially symmetric channel.
It has been found that in a small region of the parameter space the optimal decompositions are of length three.
To put it in the case we are considering, it means that the EoF of some $\rho^{BC}$ can be obtained on a 3-state ensemble.
It immediately follows that the optimal ensemble for $\rho^{B}$ in the sense of the MAE (or quantum discord) is also a 3-state one, and that the optimal measurement on qubit $A$ is a three-element POVM.
In fact, while performing numerical computation, we find the following example.
\begin{equation}
  \rho^{AB}=\left(
  \begin{array}{cccc}
    0.7 & 0 & 0 & 0.2795 \\
    0 & 0 & 0 & 0 \\
    0 & 0 & 0.15 & 0 \\
    0.2795 & 0 & 0 & 0.15
  \end{array}\right).
\end{equation}
It is a rank-three X state. If we perform von Neumann measurement on qubit $A$, the MAE can be calculated as
\begin{equation}
  \overline{S}_{\min}^{(2)}=0.295127,
\end{equation}
In this case, the optimal measurement operators are $\frac{1}{2}(\mathbbm1\pm\sigma_x)$, which give rise to the equi-entropy decomposition of $\rho^B$.

Now we perform a three-element POVM measurement on qubit $A$. The three measurement directions are given by
\begin{eqnarray*}
  \bi{k}_1=(0.929301,\,0,\,-0.369322), \\
  \bi{k}_2=(-0.929301,\,0,\,-0.369322), \\
  \bi{k}_3=(0,\,0,\,1).
\end{eqnarray*}
The corresponding probability is
\begin{equation*}
    \mathrm{Prob}_1=\mathrm{Prob}_2=0.365144,\;\;\mathrm{Prob}_3=0.269712.
\end{equation*}
Under such measurement the MAE is
\begin{equation}
  \overline{S}^{(3)}_{\min}=0.291942<\overline{S}^{(2)}_{\min}.
\end{equation}
Although the difference $\overline{S}^{(2)}_{\min}-\overline{S}^{(3)}_{\min}\approx0.003$ is small, it indicates that the three-element POVM measurement is optimal.

\section{Conclusion}

In conclusion, we consider the K-W relation from the viewpoint of quantum channel.
The quantum channel, isomorphic to the given state, determines the form of the quantum steering ellipsoid, which is a useful tool to discuss quantum discord.
In the context of K-W relation, this channel is closely related to the concurrence of the complement state.

Recently, several works are devoted to find more rigorous results of quantum discord for 2-qubit states
\cite{Lu.PhysRevA.83.012327.2011,Chen.arXiv.1102.0181,
      Girolami.PhysRevA.83.052108.2011,Galve.arXiv.1107.2005}.
Here we have shown that K-W relation, together with known results of EoF, provide us useful information for this problem.
We point out that the quantum discord of any 2-qubit rank-two state can be evaluated exactly and strictly.
In this case, the optimal measurement on one qubit is the von Neumann measurement, which will induce the equi-entropy decomposition for the reduced state of the other qubit.
The situation is more complicated for the 2-qubit states with rank larger than two.
Von Neumann measurement may not be the optimal choice.
One has to take three-element POVM measurements into consideration.

\ack

This work was supported by National Nature Science Foundation of China, the CAS, and the National Fundamental Research Program 2007CB925200.

\section*{References}

\bibliography{RankTwo}

\clearpage
\end{document}